\documentstyle[12pt,aps,prb,epsf]{revtex}
\input tcilatex

\QQQ{Language}{
American English
}

\begin{document}

\bigskip

\begin{center}
{\huge {\bf The Eventual Quintessential}}

\bigskip 

{\huge {\bf Big Collapse of the Closed}}

\bigskip 

{\huge {\bf Universe with the Present}}

\bigskip 

{\huge {\bf Accelerative Phase}}

\bigskip \bigskip
\bigskip \bigskip 

{\Large {\bf De-Hai Zhang}}\\[0pt]

\bigskip 

(e-mail: dhzhang@sun.ihep.ac.cn)\\[0pt]

Department of Physics,\\[0pt]

The Graduate School of The Chinese Academy of Sciences,\\[0pt]

P.O.Box 3908, Beijing 100039, P.R.China.\\[0pt]

\bigskip

{\bf Abstract:}
\end{center}

\baselineskip=16pt

{\hspace*{5mm}}

Whether our universe with present day acceleration can eventually collapse
is very interesting problem. We are also interesting in such problems,
whether the universe is closed? Why it is so flat? How long to expend a
period for a cycle of the universe? In this paper a simple ``slow-fast''
type of the quintessence potential is designed for the closed universe to
realize the present acceleration of our universe and its eventual big
collapse. A detail numerical simulation of the universe evolution
demonstrates that it divides the seven stages, a very rich story. It is
unexpected that the quintessential kinetic energy is dominated in the shrink
stage of the universe with very rapid velocity of the energy increasing. A
complete analytic analysis is given for each stage. Some very interesting
new problems brought by this collapse are discussed. Therefore our model
avoids naturally the future event horizon problem of the present
accelerative expanding universe and maybe realize the infinite cycles of the
universe, which supplies a mechanism to use naturally the anthropic
principle. This paper shows that the understanding on the essence of the
cosmological constant should contain a richer content.

\bigskip

\baselineskip=20pt

\section{Introduction}

The birth and termination of the universe is a riddle for mankind. An old
story is that a matter-dominated closed universe will shrink finally$^{[1]}$%
, so that the universe begins its new big bang through a bounce, and
repetition forever to realize infinite kinds of universes. However, this old
idea was challenged by a fact known recently, i.e., our present universe is
accelerated expanding$^{[2]}$. If the reason to cause the acceleration of
the universe is the real cosmological constant, then the universe will be
accelerated expanding forever, never shrink in its future, thus we can not
see the cycle of the universe. It is believed generally that the appearance
of the wisdom life needs some kind of adjustment of some physics parameters,
so that we need many worlds to realize this adjustment, that is the
anthropic principle$^{[3]}$. If the universe never shrink, we have to invoke
the concept of the chaotic inflation$^{[4]}$. The universe is inflating
eternally in a larger region, produces many many different worlds in a
fractal manner. Today this new idea is very popular in the cosmologist
community. Comparing earnestly the old and new ideas, one can find out that
the argument of the chaotic inflationary production of many universes owns
some difficulty of mathematical description, and the old idea is more simple
and clear.

We hope that our universe is accelerative in present and will shrink
finally. The advantage of this idea is that the universe maybe stays in its
infinite loop to realize many world scenario, that this model agrees with
the present decisive astronomical observations, specially the cosmic
microwave background radiation (CMBR)$^{[5]}$, and that it avoid a
difficulty about an event horizon for a perpetual accelerated universe$^{[6]}
$. In order for the universe to shrink finally there must be some negative
energy in it, for instance, a negative cosmological constant or a
positive-curvature negative-energy density. In the first case we need that
the cosmological constant changes from the present positive value to the
future negative one, thus the cosmological constant is not a true
``constant'' at all. It is better for us to accept the quintessence scheme$%
^{[7]}$, which has a good basis of the field theory. Anyhow we hope to avoid
a ``true cosmological constant'' further, no matter what it is positive or
negative, which is a big obstacle to construct a string (M-) theory$^{[8]}$.
Therefore we take the vacuum of the quintessence, i.e., the minimum of its
potential, to be the zero point of the energy, that is the true cosmological
constant is zero absolutely for some unknown reason$^{[9]}$. The force to
shrink the universe is remained only as the positive-curvature
negative-energy density, i.e., the universe is closed. In spite of the
universe is closed, we shall show in the section 10 that a reasonable
assumption from inflation of the early universe gives a very small
non-flatness of the universe, which is unobservable at present.

It is a very interesting problem how to shrink finally for a closed
universe, which has acceleration at present caused by a special
quintessence. It should be known that for a generic quintessence, for
instance with a pure inverse power law potential or a pure exponential one,
it can not arrive the above goal. That the universe is accelerative at
present and will shrink in future is to require that quintessence is a
``(early) slow - (late) fast (rolling)'' type. However, the inverse power
quintessence is a ``fast-slow'' type, and the exponential quintessence is a
``slow-slow'' or ``fast-fast'' ones. So that we must design a reasonable
``slow-fast'' quintessence. In this paper we do it in the section 2. This
potential must have as simpler form and fewer parameters as possible. This
special potential is called by us as ``Niagara'' one, since its
configuration is very flat for its negative field value and falls very
rapidly to the zero for its positive field value, and the quintessence rolls
down from a high potential hill to a low valley without boundary.

Whether there is some new story about the evolution of the universe with our
``Niagara potential''? Yes, some rich and wonderful scenario appears
unexpectedly. As a typical process from the Big Bang Neucleonsythesis (BBN)
to the big collapse of the universe, the evolution may be divided into seven
stages, as described in the section 3. The first and second stages are the
radiation and matter dominated epochs respectively, which are familiar for
us. The third stage is the quintessence potential energy dominated, and the
universe stays in its accelerative phase, in which the quintessence rolls
slowly with its negative field value (see analysis in the sections 4-5).
However, the quintessence rolls inevitably to its positive field value. When
it passes through its turning point, the quintessence potential energy
decreases suddenly, thus the quintessence kinetic energy will be dominated,
and this is the fourth stage of the universe evolution, which is decelerated
(see analysis in the section 6). Since the evolution law of the quintessence
kinetic energy is faster than matter, when the cosmic scale factor expand to
some value the quintessence kinetic energy will be approximately equal to
the matter energy, the universe comes into its fifth phase, i.e., the
``matter leading attractor'' one, since its evolution law is like of matter
and this new term will be explained in the section 7. Moreover, the
evolution law of the matter energy is faster than the curvature energy, at
some time the universe is dominated by the curvature energy, i.e., the sixth
phase of the evolution (see analysis in the section 8). When the curvature
negative energy cancels just all of other energies, the universe stop its
expansion, and begin to shrink, i.e., the seventh phase. We note
interestedly that when the universe begin to shrink, the quintessence
kinetic energy, the matter one and the curvature one are all approximately
equal. As the universe shrinks, the quintessence kinetic energy increase
more quickly than other form energies, soon after it the quintessence
kinetic energy is dominated in the collapse universe, therefore the seventh
phase of the universe evolution is the ``quintessence kinetic energy
collapse'', it is very different from the old concept of the matter collapse
(see analysis in the section 9). This will bring many interesting problems
to deserve further investigation (see the section 11).

It is worth to be emphasized that it is very necessary to finish a numerical
simulation in computer for a whole evolution process of the universe,
otherwise we can not supply a time-scale concept for each stage and a
diminishable scenario of the universe. We noted often that many seeming
feasible models are actually impracticable, specially to obtain a suitable
accelerative period and a reasonable age for the present universe. We give a
detail result of the simulation of this seven stage evolution in this paper.
We need not only the detail simulation (in the section 3), but also an
analytic analysis for its essence in a manner of the quantitative estimation
(in the sections 4-9).

The model parameters taken by us is only a special point of the wide
parameter space in our model, however it is typical for an understanding of
the essence of the evolution. To explore the various behavior of the whole
parameter space is not the task of this paper. We only hope to express an
idea about the present accelerated expanding universe still has a chance to
shrink eventually. This opens a way to realize infinitely many universes.
This kind of many universe is more simple and easy to be described in order
to serve the anthropic principle. Finally we make a simple conclusion in the
section 12, specially about interesting problems worth to be investigated
further.

\section{The potential of the quintessence}

We have announced that it is necessary to replace a true cosmological
constant with the quintessence field $\phi $. The quintessence potential we
shall take in our model has the following form 
\begin{equation}
\label{z010918a}V(\phi )=\frac{M_0^4}{1+\exp (\lambda \phi /M_p)}.
\end{equation}
When $\phi \rightarrow -\infty $, the potential $V(\phi )$ approaches to $%
M_0^4$, which emulates realistically a true cosmological constant, we take $%
M_0=0.95\times 10^{-30}M_p$, here $M_p=(8\pi G)^{-1/2}$ $=2.4\times 10^{18}$%
GeV is the Planck energy scale. In other hand when $\phi $ is larger than
zero, the potential $V(\phi )$ declines exponentially and approaches to zero
quickly. The slow roll parameter of this potential is 
\begin{equation}
\label{z010918b}\epsilon \equiv \frac 12M_p^2\frac{V^{\prime 2}}{V^2}=\frac{%
\lambda ^2/2}{(1+e^{-\lambda \phi /M_p})^2}.
\end{equation}
The parameter $\lambda $ will control the velocity of rolling. This
potential is ``slow-slow'' type for $\lambda <\sqrt{2}$ and ``slow-fast''
one for $\lambda >\sqrt{2}$. In our model we take $\lambda =2$ as a typical
analysis. Due to the shape of this potential likes the Niagara fall, we call
it as ``Niagara potential''. The field value $\phi =0.44M_p$ is the turning
point between the slow and fast rolling. When $\phi <0.44M_p$ the
quintessence is slow roll with $\epsilon <1$ and when $\phi >0.44M_p$ it
rolls fast with $\epsilon >1$. The model with $\lambda =2$ is a typical
``slow-fast'' potential.

The potential of the inverse power law $M_0^{4+\alpha }\phi ^{-\alpha }$ of
the quintessence is the ``fast-slow'' type which can realize the interesting
tracker solution$^{[10]}$. The pure exponential potential $\exp (\lambda
\phi /M_p)$ is ``slow-slow'' one for $\lambda <\sqrt{2}$ and ``fast-fast''
for $\lambda >\sqrt{2}$. Of course we can design a more compound
``fast-slow-fast'' potential to realize both tracker and shrinkage, but this
is not a purpose of this paper.

\section{The simulation of the universe evolution}

The whole evolution obeys the following famous equation, which was
researched for thousands times in the literatures, 
\begin{equation}
\label{z010918c}H^2=\frac 1{3M_p^2}(\rho _\phi +\rho _m+\rho _r+\rho _c),
\end{equation}
\begin{equation}
\label{z010918d}\ddot \phi +3H\dot \phi +V^{\prime }=0,
\end{equation}
where $H=\dot a/a$ is the Hubble parameter and $a$ is the cosmic scale
factor, $t$ is the cosmic time, $\rho _\phi =\rho _k+\rho _v$ is the
quintessence (total) energy density, in it $\rho _k$$=\dot \phi ^2/2$ is the
quintessence kinetic one, $\rho _v=$$V(\phi )$ is the quintessence potential
one, $\rho _m=\rho _{m0}(a/a_0)^{-3}$ is the matter one, $\rho _r=\rho
_{r0}(a/a_0)^{-4}$ is the radiation one, and $\rho _c=\rho _{c0}(a/a_0)^{-2}$
is the curvature one which is negative for a closed universe. We noted $\rho
_{m0}$, $\rho _{r0}$, $\rho _{c0}$ are the parameters of the model together
with $\lambda $ and $M_0$, total five. A set of initial condition, $a_0$, $%
\phi _0$, $\dot \phi _0$ at the initial time $t_0$, determines uniquely the
whole evolution from the BBN to final constriction of the universe. As a
typical analysis, we take the following parameters, $t_0=1$sec$=0.4\times
10^{43}M_p^{-1}$ for BBN time, $\rho _{r0}=2.6\times 10^{-86}M_p^4$, $\rho
_{m0}=2.2\times 10^{-92}M_p^4$ for an almost realistic universe model. We
shall see that it gives a set of the suitable parameters of the present
universe. The initial condition is taken as $\phi _0=-2$ in order to have an
enough slow rolling, and $\dot \phi _0=0$, i.e., no roll at beginning, we
shall see that this assumption is reasonable. As for the curvature energy,
we take $\rho _{c0}=-10^{-182}M_p^4$, we shall explain why this value should
have such a small magnitude orders. $a_0$ is arbitrary unit for a while, we
shall determine which value it should have. If we know the values of $a$, $%
\phi $ and $\dot \phi $ (it is determined by function $\phi (t)$) at an
arbitrary time $t$, we can calculate the values of all other 15 quantities
at this time, such as $H$, $\rho _m$, $\rho _r$, $\rho _c$, $\rho _k$, $\rho
_v$, the various density ratios $\Omega _j=\rho _j/\rho _{crit}$, where the
critical density is $\rho _{crit}=3M_p^2H^2$, the quintessence state
equation $w_\phi =(\rho _k-\rho _v)/(\rho _k+\rho _v)$ and the acceleration
parameter $q=a\ddot a/\dot a^2$ in according to the formulae just given, as
well as 
\begin{equation}
\label{z010929a}\ddot a=\frac a{3M_p^2}(\rho _v-2\rho _k-\rho _m/2-\rho _r).
\end{equation}
The key is to obtain $a$, $\phi $ and $\dot \phi $ as the functions of time.
Remember that this principle can guide our later analytic analysis.

It is impossible to obtain the exact solution of the whole process of this
set of equations Eqs.(\ref{z010918c}) and (\ref{z010918d}). We do a fine
numerical simulation. The result is unexpectedly rich. The evolution divides
seven stages, all listed numbers are obtained by computer calculation, the
purpose that some numbers preserve their high precision form is for
convenience in order to compare with the later analytic analysis.

The result of the simulation is mainly given in the Figs.1-4, the evolutions
of $q$, $a$, $\phi $ and $\dot \phi $. Some quantities have too large
magnitude orders, so that we have to use a double logarithmic coordinates.
The seven stages of the evolution can clearly be seen in the figure of $q$
value, as a feature quantity of the different stage. It is a pity that we
are not able to put the wonderful evolution figures of all other 14
quantities in this paper for lack of a page space. Due to the continuity of
the evolution process, the turning points among the different stages have to
be chosen carefully, even subtly, otherwise it is not easy to obtain the
good results in order to compare with the analytic analysis.

1) The radiation dominated epoch, from $t_0$ to $t_1=0.6\times
10^{55}M_p^{-1}$, where time $t_1$ is defined as $\rho _{r1}=\rho _{m1}$,
i.e., the radiation-matter equal. The cosmic scale factor increases to $%
a_1=1.2\times 10^6a_0$. The quintessence field value is almost invariant $%
\phi _1=-2M_p$. The quintessence velocity is $\dot \phi _1=10^{-67.2}M_p^2$.
The acceleration parameter takes its value $q_{(0\rightarrow 1)}\simeq -1$
and $q_1=-0.75$, i.e., the universe is decelerated expanding in this stage.
The quintessence state equation is almost $w_{\phi (0\rightarrow 1)}\simeq
-1 $.

2) The matter dominated epoch, from $t_1$ to $t_2=0.86\times 10^{60}M_p^{-1}$%
, where time $t_2$ is defined as $\rho _{m2}=2\rho _{v2}$, i.e., the
matter-``cosmological constant'' approximately equal. The reason why we
choose a coefficient $2$ is that at this time the universe comes just into
the accelerative phase, $q_2=0.005$. Before this point $q_{(1\rightarrow
2)}\simeq -0.5$. The quintessence has $\phi _2=-1.997M_p$, $\dot \phi
_2=10^{-62.1}M_p^2$, $w_{\phi 2}=-0.9999$. The cosmic scale factor increases
to $a_2=2.4\times 10^9a_0$.

3) The ``cosmological constant'' dominated epoch, from $t_2$ to $%
t_3=3.2\times 10^{61}M_p^{-1}$, where time $t_3$ is defined as $\phi
(t_3)\simeq $$0$, i.e., the ending of the quintessence slow rolling. In this
epoch the universe stays in the inflationary phase $q_{(2\rightarrow
3)}\simeq 1$ and $q_3=0.676$. The quintessence has $\phi _3=0.026M_p$, $\dot
\phi _2=10^{-60.5}M_p^2$, $w_{\phi 3}=-0.784$ and $w_{\phi (2\rightarrow
3)}\simeq -1$. The cosmic scale factor increases to $a_3=1.7\times
10^{16}a_0 $. Here the result of the point $3$ is new.

In particular, this epoch contains a special point, our present universe,
denoted as ``$u$''. We have $t_u=1.56\times 10^{60}M_p^{-1}=13$Gyr, i.e.,
the universe age. Maybe you noted that this number solves the crisis of the
universe age$^{[11]}$ due to the existence of the pseudo ``cosmological
constant''. The time point $t_u$ is defined as $\rho _v$$(t_u)=(7/3)$$\rho _m
$$(t_u)$, i.e., $\Omega _{mu}=0.3$ and $\Omega _{vu}=0.7$. The present
universe entered in the accelerative phase not long ago, $q_u=0.55$. In this
point we have $\rho $$_{ru}=10^{-124}M_p^4$ which is just the energy density
of the present CMBR with $T_u=2.726^oK$. The Hubble constant is $H_u=100h$km$%
\cdot $sec$\cdot $Mpc$^{-1}$ $=0.62\times 10^{-60}M_p$ , then $h=0.7$.
Through this data we can see the model parameters are chosen reasonably. The
scale factor of the present universe is $a_u=4\times 10^9a_0$.

4) The following results are new, you must note the jumping changes of
various parameters. The fourth stage is the ``kinetic energy leading
attractor'' epoch, from $t_3$ to $t_4=1.6\times 10^{101}M_p^{-1}$, where
time $t_4$ is defined as $\rho _m$$(t_4)=$$0.3\rho _v$$(t_4)$, i.e., the
matter-quintessence approximated equal, why such choice will be explained in
later analysis. After $t_3$ the universe turns suddenly into the decelerated
phase, $q_{(3\rightarrow 4)}\simeq -1$ and $q_4=-0.84$. The quintessence has 
$\phi _4=95.4M_p$, $\dot \phi _4=10^{-101.2}M_p^2$, $w_{\phi 4}=0.25$ and $%
w_{\phi (3\rightarrow 4)}\simeq 1/3$. The cosmic scale factor increase to $%
a_4=1.8\times 10^{37}a_0$.

5) The ``matter leading attractor'' epoch, from $t_4$ to $t_5=0.28\times
10^{182}M_p^{-1}$, where time $t_5$ is defined as $\Omega $$_c(t_5)\simeq -1$%
, i.e., the curvature dominated. The scale factor of the universe is $%
a_5=3.1\times 10^{90}a_0$. During this epoch the universe is in the
decelerated phase, $q_{(4\rightarrow 5)}\simeq -0.5$ and $q_5=-1.55$. The
quintessence has $\phi _5=280M_p$, $\dot \phi _5=10^{-181.4}M_p^2$, $w_{\phi
5}=0.26$ and $w_{\phi (4\rightarrow 5)}\simeq 0$. The density ratios of the
simulation are $\Omega _{k5}=0.823$, $\Omega _{m5}=0.770$, $\Omega
_{v5}=0.482$ and $\Omega _{c5}=-1.075$.

6) The ``curvature dominated'' epoch, from $t_5$ to $t_6=0.85\times
10^{182}M_p^{-1}$, where time $t_6$ is defined as $H$$(t_6)=$$0$, i.e., the
absolute value of the curvature energy equals the sum of all other energies,
called it as the ``balanced point'', and the scale factor of the universe
arrive its maximum value $a_6=4.5\times 10^{90}a_0$. During this epoch the
universe is still in the decelerated phase $q_6\rightarrow -\infty $. The
quintessence has $\phi _6=281.5M_p$, $\dot \phi _6=10^{-181.67}M_p^2$, $%
w_{\phi 6}=0.805$. In fact the computer value of $H_6$ is not true zero due
to lack of the calculation precision. In this point we have the interesting
proportions $\rho $$_{k6}/|\rho _{c6}|=0.461$, $\rho $$_{m6}/|\rho
_{c6}|=0.489$ and $\rho $$_{v6}/|\rho _{c6}|=0.050$ for various energies. We
avoid to use $\Omega _{j6}$ here due to their infinity at the balance point.

7) The big shrink epoch, from $t_6$ to $t_8\simeq 1.74t_6$ for a rough
estimation, where time $t_8$ is defined as $a_8\simeq 0$, i.e., the cosmic
scale factor shrinks suddenly to almost zero! However it is very difficult
to look for this point $t_8$ in computer, instead of we choose another
neighboring point $t_7=1.47\times 10^{182}M_p^{-1}=1.73t_6$ to study the
behavior of the cosmic constriction. At this point $a_7=0.27\times
10^{90}a_0\simeq a_6/17$, indeed the universe is shrinking. We have $\phi
_7=289.1M_p$, $\dot \phi _7=10^{-178}M_p^2\simeq 4700\dot \phi _6$, $w_{\phi
7}\simeq 1$, i.e., the quintessence kinetic energy becomes large rapidly.
The computer calculation gives $\Omega _{k7}=0.99981$ and $\Omega
_{m7}=0.00021$, $\Omega _{c7}=-0.00003$, the rests are almost zero, i.e., it
is a ``quintessence kinetic energy dominated shrink''. This is a very
interesting new result obtained by this paper! We shall illustrate its
important meaning in later.

Maybe one notes that many terms and scenario have not been illustrated in
above text. All things can be done in the following detail analysis, we need
patience. This numerical simulation demonstrates a vivid story of the
universe evolution. In order to make one believe this simulation, we must do
some convincing analytic analysis, and look for the profound relations
between various data. We must answer the question such as what is the
features of the balanced point, the essence and inevitability of the shrink,
and the reason of the closed universe with sufficient flatness.

\section{The evolution of the quintessence in the radiation or matter
dominated}

At the first and second epochs we meet a problem how the quintessence
evolves in the radiation or matter dominated environment, $\rho _n\gg \rho
_\phi \gg |\rho _c|$, where $\rho _n=\rho _{n0}a^{-n}$ with $n=3$ for matter
and $n=4$ for radiation. Suppose the quintessence is in slow roll process,
i.e., $\phi <0$ and $\ddot \phi \simeq 0$. We have $a\propto t^{2/n}$ and $%
H=2/(nt)$, and from Eq.(\ref{z010918b}), 
\begin{equation}
\label{z010918e}3\cdot \frac 2{nt}\dot \phi -\frac \lambda {M_p}\cdot \frac
1{1+e^{-\lambda \phi /M_p}}\cdot \frac{M_0^4}{1+e^{\lambda \phi /M_p}}=0.
\end{equation}
Integrating it we obtain the time $t_e$ at which the quintessence ends its
slow rolling from the initial value $\phi _0$ to the final one $\phi
_e\simeq 0$ (here index ``$e$'' means ``end''), 
\begin{equation}
\label{z010919b}t_{e(1,2)}\simeq M_0^{-2}\sqrt{\frac{12}n}\frac{M_p}\lambda
e^{-\lambda \phi _0/(2M_p)}.
\end{equation}
It doesn't matter for this estimation to use $\phi _e\simeq 0$ rather $\phi
_e\simeq 0.44M_p$. The time $t_e$ must be larger than the present universe
age $t_u$. Otherwise, there is not an enough time to allow a phase in which
the cosmological constant be dominated, since the quintessence already rolls
fast. Once the quintessence comes into fast rolling, the evolutionary law of
the quintessence will be changed, does not obey this formula rather the
later Eq.(\ref{z010919e}). From it we know that it is safe to take $\phi
_0=-2$, since $t_e=4.4t_u$. From Eq.(\ref{z010918c}) we also get the
quintessence $\phi $ evolution of slow rolling in the matter or radiation
dominated epochs 
\begin{equation}
\label{z010919a}\phi \simeq -\lambda ^{-1}M_p\ln (e^{-\lambda \phi
_b/M_p}-n\lambda ^2M_p^{-2}M_0^4t^2/12)\equiv \phi _{-}(t),
\end{equation}
moreover which is also valid in the whole epoch of the quintessential slow
rolling, even if the quintessence dominated, i.e., $t_0<t$ $<t_3$. Using Eq.(%
\ref{z010919a}) we can estimate $\tilde \phi _1=-2(1-O(10^{-13}))M_p$, $%
\tilde \phi _2=-1.997M_p$ and $\stackrel{\sim }{\dot \phi }_1\simeq \,%
\stackrel{\sim }{\dot \phi }_2\simeq 10^{-62}M_p^2$, coincides with the
simulation value. Hereafter the wave ``$\sim $'' means the analytic
estimated values. the estimation of the theoretical values of $a_1$ and $a_2$
is elementary and familiar, and the values of other various quantities at
the time point $2$ are all get in according to the mentioned principle, thus
we omitted these calculations.

\section{The slow rolling quintessence dominated}

At the third epoch we meet a problem how the quintessence evolves in the
quintessence dominated, $\rho _\phi \gg \rho _n\gg |\rho _c|$, and $\epsilon
<1$. This problem has been solved partly in the last section, here we
concern the scale factor. This is a typical inflationary process $a\propto
e^{Ht}$ with $H\simeq $const., in which the quintessence rolls slowly from $%
\phi _b\simeq \phi _2$ to $\phi _e\stackunder{\sim }{<}\,0.44M_p$ (The index
``$b$'' means ``begin'' in this paper), we can use its standard formulas,
i.e., the inflationary e-folding $N$ is given by 
\begin{equation}
\label{z010919c}N=\ln \frac a{a_0}=\int_{\phi _b}^{\phi _e}M_p^{-2}(-\frac
V{V^{\prime }})d\phi .
\end{equation}
Using it we have 
\begin{equation}
\label{z010919d}N=\lambda ^{-1}M_p^{-1}(\phi _e-\dfrac{M_p}\lambda
e^{-\lambda \phi _e/M_p})+\lambda ^{-1}M_p^{-1}(-\phi _b+\dfrac{M_p}\lambda
e^{-\lambda \phi _b/M_p}),
\end{equation}
and the time at which the quintessence slow rolling is terminated, 
\begin{equation}
\label{z010919e}t_{e(3)}\simeq \sqrt{3}\lambda ^{-2}M_pM_0^{-2}e^{-\lambda
\phi _b/M_p},
\end{equation}
which must be larger than the present universe age $t_u$.

However we are not able to analysis the data of the time point $3$ for a
while since it involves the next stage evolution. After we finish the
analytic analysis of the next stage, we shall return back to do a
theoretical estimation of the data of the third cosmic stage.

\section{The quintessence kinetic attractor}

A question we face in the fourth stage of the universe evolution is the fast
rolling quintessence dominated, i.e., from the time points $3$ to $4$, $\rho
_\phi \gg \rho _n\gg |\rho _c|$, with $\phi >M_p$ and $\epsilon >1$. It is
unexpected that its solution is very simple 
\begin{equation}
\label{z010920a}\phi =2\lambda ^{-1}M_p\ln (\lambda M_p^{-1}M_0^2t)\equiv
\phi _{+}(t),
\end{equation}
which can be easily checked by putting it into Eqs.(\ref{z010918c}) and (\ref
{z010918d}), furthermore only this formula is still valid in the fifth and
sixth stages, other formulae will different. Its character is that the
quintessence state equation is $w_\phi =\lambda ^2/3-1$ and the evolution of
the cosmic scale factor is $a\propto t^{2\lambda ^{-2}}$, the acceleration
parameter $q=1-\lambda ^2/2$. When we take $\lambda =2$ the evolution of
this quintessence is just like radiation $w_r=1/3$, but it is only a
fortuity. In this case its kinetic energy is twice of its potential, $\rho
_k=2\rho _v$. It is seen clearly that there is not the decelerating of the
universe for the ``Niagara'' quintessence with $\lambda <\sqrt{2}$.

Now a careful observation is needed for the slow-fast turning point. When
the quintessence rolls slowly it obeys the evolution Eq.(\ref{z010919a}),
and when it rolls fast it obeys the evolution Eq.(\ref{z010920a}). Observing
the neighbor of the point $3$ of the simulation curve in the Fig.3, which
has a subtle transition, we have a good approximation about this point. We
suppose that $\phi =\phi _{-}$ for $t<\tilde t_3$ and $\phi =\phi _{+}$ for $%
t>\tilde t_3$ and $\phi _{-}=\phi _{+}$ for $t=\tilde t_3$. These two curves
intersect at the time point $\tilde t_3$, we can obtain the cross point $%
\tilde t_3=2.7\times 10^{61}M_p^{-1}$ and $\phi _{-}($ $\tilde t_3)$ $=$ $%
\phi _{+}($ $\tilde t_3)$ $=3.888M_p$ by solving union equations. This point
is near the slow-fast turning point, $\phi =0.44M_p$, therefore the
inflationary e-folding formula Eq.(\ref{z010919d}) is still valid, we take
the slow rolling ending point as $\phi _e=$$\phi _{-}($ $\tilde t_3)$ and
the beginning point is $\phi _b\simeq -2M_p$ then we have from Eq.(\ref
{z010919d}) $\tilde N=16.5$, then $\tilde a_3=1.5\times 10^7\tilde a_2$. Due
to the delicacy of the transition point $3$, it is reasonable to take the
theoretical value of $\tilde \phi _3$ as almost zero (therefore different
from the value $\phi _{-}($ $\tilde t_3)$, a skill for us), and obtain $%
\stackrel{\sim }{\dot \phi }_3\simeq 10^{-60}M_p$ from Eq.(\ref{z010920a}).
These estimations are good coincident with the simulation data.

\section{ Quintessence-matter attractor}

A question we face in the fifth stage of the universe evolution is the fast
rolling quintessence domination with the matter leading, i.e., from the time
points $4$ to $5$, $\rho _\phi \simeq \rho _m\gg \rho _r\gg |\rho _c|$.
Putting $a\propto t^{2/n}$ and $H=2/(nt)$ into Eqs.(\ref{z010918c}) and (\ref
{z010918d}), we can find out that 
\begin{equation}
\label{z010920b}\Omega _m=1-\frac n{\lambda ^2}=\frac 14,\qquad \Omega _k=
\frac{n^2}{6\lambda ^2}=\frac 38,\qquad \Omega _v=\frac n{\lambda ^2}-\frac{%
n^2}{6\lambda ^2}=\frac 38.
\end{equation}
This the attractor solution is given by Ref.[12], we call it as the ``matter
leading attractor'' if $n=3$ and the ``radiation leading'' one for $n=4$ due
to their evolution behavior. Now we can understand why we choose the time
point $4$ as $\rho _m$$(t_4)=$$0.3\rho _v$$(t_4)$ since the matter density
increases from $\Omega _m/\Omega _v\simeq 0$ to $\Omega _m/\Omega _v\simeq
2/3$ when the universe comes from the quintessence leading phase to the
matter leading attractor, we only take a middle point $\Omega _m/\Omega
_v\simeq 0.3$ as this turning point of the simulation.

However when we estimate other theoretical values, it is impossible to adopt
an exact manner. The quintessence has its energy density about $\tilde \rho
_{v3}\simeq $$10^{-120}M_p^4$ at the ending point $3$ of the slow rolling,
and the matter density is about $\tilde \rho _{m3}\simeq 10^{-141}M_p^4$ at
this point. The evolution of the quintessence energy is like of the
radiation, $\rho _\phi $$\propto a^{-4}$, and in other hand $\rho _m$$%
\propto a^{-3}$ for matter. When the cosmic scale factor expands to $\tilde
a_4/\tilde a_3=\tilde \rho _{v3}/\tilde \rho _{m3}=10^{21}$, the
quintessence energy will decrease faster and is approximated equal to the
matter energy, the universe enters from the quintessence kinetic dominated
phase into the matter leading attractor phase. The time is $\tilde
t_4/\tilde t_3\simeq (\tilde a_4/\tilde a_3)^2=10^{42}$. Then it is easy to
obtain $\tilde \rho _{m4}\simeq 10^{-204}M_p^4$ and $\tilde \rho _{c4}\simeq
10^{-257}M_p^4$. The evolution of $\phi $ is still Eq.(\ref{z010920a}), one
gets $\tilde \phi _4=99M_p$ and $\stackrel{\sim }{\dot \phi }_4\simeq
10^{-101}M_p^2$. All estimation is confirmed with the simulation well.

The evolution of the matter leading attractor is like of matter, $\rho _\phi
=3\rho _m\propto a^{-3}$, but the evolution of the curvature energy is
slower, $\rho _c\propto a^{-2}$, thus when the cosmic scale factor expands $%
\tilde a_5/\tilde a_4=\tilde \rho _{m4}/|\tilde \rho _{c4}|=10^{53}$, the
curvature energy will dominated in the universe. The time at which the
universe begins its curvature domination is $\tilde t_5/\tilde t_4\simeq
(\tilde a_5/\tilde a_4)^{3/2}=10^{79}$. We choose $\Omega _{c5}\simeq -1$ as
the definition of the turning point $t_5$ in the simulation. For same reason
one gets $\tilde \phi _5=282M_p$ and $\stackrel{\sim }{\dot \phi }_5\simeq
10^{-182}M_p^2$. Again all estimations are coincident with the simulation
well.

\section{The stop of the expansion and the start of the shrink of the
universe}

In the curvature energy dominated universe the curvature negative-energy
will cancel strongly with the matter and quintessence energies, the Hubble
parameter of the universe will become smaller and smaller. Ultimately, the
Hubble parameter will become zero at the time point $t_6$, and the universe
stops its expansion, $\dot a_6=0$, which is the balanced point. The time $%
t_6 $ is not far from the time $t_5$, but very near, so that the values of
the various parameters are not changed too much, $a_6\simeq a_5$, $\dot \phi 
$$_6\simeq \dot \phi _5$, $t_6\simeq t_5$. We define the time difference $%
\tau \equiv $$t-t_6$. Through the observation of the various curves of
neighbor of the balanced point of the simulation, we find out the following
expression is a good approximated solution for a narrow region $|$$\tau
|<0.01t_6$, 
\begin{equation}
\label{z010920b}\phi =\phi _6+\dot \phi _6\tau +\frac 12\ddot \phi _6\tau
^2,\qquad a=a_6-\frac 12\ddot a_6\tau ^2. 
\end{equation}
Putting it into Eqs.(\ref{z010918c}) and (\ref{z010918d}), we have 
\begin{equation}
\label{z010920c}\dot \phi _6^2=2(\,|\rho _{c0}|a_6^{-2}-M_0^4e^{-\lambda
\phi _6/M_p}-\rho _{m0}a_6^{-3}), 
\end{equation}
\begin{equation}
\label{z010920d}\ddot \phi _6=\lambda M_p^{-1}M_0^4e^{-\lambda \phi _6/M_p}, 
\end{equation}
\begin{equation}
\label{z010920e}\ddot a_6=M_p^{-2}(1+3\lambda ^{-2})^{-1}(\frac 23|\rho
_{c0}|a_6^{-1}-\frac 12\rho _{m0}a_6^{-2}). 
\end{equation}
One can check easily that the theoretical estimation data obtained by using
theses formulas are almost same with the simulation data. Let us look at the
Hubbel constant, $H=-a_6^{-1}\ddot a_6\tau $, we have $H>0$ when $\tau <0$
and $H<0$ when $\tau >0$. The universe comes from the expansion to the
shrinkage. It is very important that the sign of the Hubble parameter is
changed, which turns the direction of the universe evolution.

\section{The big shrink of the quintessence kinetic energy domination}

We have to note that at the balanced point, where $a_6$ is at a maximum of
the cosmic scale factor, the proportion of the quintessence kinetic energy
is not small, $\Omega _k/\Omega _m=0.94$ (note that $\Omega _k,\Omega
_m\rightarrow \infty $ at this time point). Once the universe comes into its
contractive phase, the proportion of the quintessence kinetic energy will
increase rapidly. A good approximation for this phase is to set $\rho
_v=\rho _m=\rho _r=\rho _c=0$. This is a typical quintessence kinetic
dominated model, the unique distinction is the sign of the Hubble parameter,
which is now negative! But its evolution law is same, $\rho _k=$$\dot \phi
^2/2\propto a^{-6}$, when the universe shrinks, the quintessence kinetic
energy increases faster than any other energies. The just same behavior in
the early expanding universe is used to construct a tracker solution$^{[10]}$%
. The quintessence begins its speedy running at the very flat potential. Its
running energy make universe to shrink more rapidly, and then the kinetic
energy is higher. Using this simplification we obtain easily its solution
for $\tau >t_7-t_6$ from Eqs.(\ref{z010918c}) and (\ref{z010918d}), 
\begin{equation}
\label{z010920f}\phi =\phi _6-\sqrt{6}M_p\ln (a/a_6),\qquad a=a_6(1-\sqrt{3/2%
}\dot \phi _6M_p^{-1}\tau )^{1/3}. 
\end{equation}
If any unforeseen does not happen, the universe will shrink to zero size $%
a_8=0$ during the time interval $\tau _8=\sqrt{2/3}\dot \phi
_6^{-1}M_p\simeq 2t_6$ from (\ref{z010920f}). In a view of the exponential
time the universe shrinks to zero size almost suddenly. It is very different
from the shrinkage of the matter dominated universe familiar for us. The
time $t_8=t_7+\tau _8\simeq 10^{182}M_p^{-1}\simeq 10^{122}t_u$ is the life
of a cycle of the universe in our model, which is an unimaginable very long
time, but it is important that it is finite.

\section{The some conjunction about a small initial curvature energy}

After rearrangement of the catenulate relations among the scale factors of
the various evolution stages, we actually obtain a simple expression about
the maximum cosmic scale factor for a cycle of the universe 
\begin{equation}
\label{z010920h}a_6=\dfrac{\rho _{v2}}{|\rho _{c2}|}a_2,
\end{equation}
i.e., it is mainly determined by the initial curvature energy. The formula
is valid only if the $a_6$ is larger than $a_3$, since only if the
quintessence ends its slow rolling we are able to use this formula
correctly. A focus question is why the curvature energy is so small in the
initial condition? It involves a quite different stage of the universe
evolution, i.e., the creation and inflation of the universe. We only can do
a very simple discussions about this important problem since the major issue
of this paper is about the present acceleration and the future constriction
of the universe.

Suppose that we have a inflation field $\Phi $ which inflationary potential
is the typical chaotic one$^{[3]}$, i.e., $V(\Phi )=M_i^2\Phi ^2/2$ (here
the index ``$i$'' means ``inflation''). Suppose that the inflaton begins its
inflation at the initial field value $\Phi _b=25M_p$ and ends its inflation
at the final value $\Phi _e=\sqrt{2}M_p$ with $\epsilon (\Phi _e)=1$ in our
model. We see that this $\Phi _b$ is very far smaller than the critical
value $\Phi _{chaotic}\stackunder{\sim }{>}\,300M_p$ at which the universe
star its true chaotic inflation behavior$^{[13]}$. The fluctuation $\delta
^2\simeq 10^{-10}$ of the CMBR observed by us will add a constraint on the
important parameter of the inflationary model, i.e., the inflaton mass, then 
$M_i^2\simeq 10^{-11}M_p^2$ is obtained from this constraint. Therefore the
inflationary e-folding is $N=\Phi _b^2M_p^{-2}/4\simeq 156$, and we have $%
a_e=10^{68}a_b$, where $a_b$ is the size of the universe at its quantum
creation. In spite of this number is very large, it is actually very smaller
than of the typical chaotic inflation. It is a good idea that the original
universe comes from an four-dimensional instanton $S^4$ with the radii $a_b$
in the Euclidean space (or to multiply directly another unvarying compact
manifold in an extra dimension space to fit the need of the string (M-)
theory$^{[14]}$). Only the closed universe ($K_c=1$) can be transformed from
this instanton, and can supply a barrier to realize its quantum tunneling,
which produces a model with the initial condition $H_b=0$ and $\dot \Phi _b=0
$. This implies that the negative value of the curvature energy $-\rho
_c=K_ca^{-2}$ must equal to the inflaton field energy in Eq.(\ref{z010918c})
at the born time of the universe$^{[15]}$ 
\begin{equation}
\label{z010921a}H_b^2+\frac{K_c}{a_b^2}=\frac 1{3M_p^2}(\frac 12M_i^2\Phi
_b^2).
\end{equation}
Thus the we obtain $a_b\simeq 3\times 10^4M_p^{-1}$ and $\rho
_{cb}=-a_b^2\simeq -10^{-9}M_p^4$, which is negative for the closed
universe. At the ending of the inflation of the inflaton the curvature
energy becomes $\rho _{ce}\simeq -10^{-145}M_p^4$ and the inflation field
energy transform into the radiation energy $\rho _{re}=\rho _{\Phi
e}=M_i^2\Phi _e^2/2=10^{-11}M_p^4$, this progress is the ``reheating'',
which temperature is about $T_e\simeq \rho _{re}^{1/4}\simeq 10^{-3}M_p$.
The BBN temperature is about $T_0=1$MeV$\simeq $$10^{-21}M_p$. From the
reheating to BBN, the universe expands by $a_0/a_e=T_e/T_0=10^{19}$,
therefore we have the parameters at BBN time $\rho _{c0}\simeq
-10^{-182}M_p^4$ and $\rho _{r0}\simeq 10^{-86}M_p^4$, which has been
adopted by us in the section 3. Even if the reheating temperature is lower
than $T_e$ just given, we can consider a period of the matter dominated
expansion of the coherent condensate state before the reheating and obtain a
correct expanding ratio, here $a_0$ is only a naive estimation. The
electro-weak interaction produces the dark matter, the baryogenesis produces
the asymmetry of the baryonic matter, so that it is reasonable that the
matter density is lower than the radiation one for many magnitude orders $%
\rho _{m0}\simeq 10^{-92}M_p^4$. If $\Phi _b$ is only small like $16M_p$,
the present scale factor has the same size of the magnitude order with the
present event horizon of our universe. Therefore at least the initial value
of the inflaton must be larger than $16M_p$. Then we see that the initial
value $25M_p$ of the inflaton set by us is not too higher than a necessary
value for the realistic universe and far smaller than the chaotic critical
value. Even so this has brought a very small curvature energy density, and
the concrete value of $\rho _{c0}$ taken here is only as an example.

We would like say some words about the closed universe. If the universe was
born by a quantum process, it is relatively easy to image that a finite size
thing was produced from nothing (for example, instanton), but it is
relatively difficult to image an infinite size thing was produced from
nothing, in spite of this is a mysterious ``quantum process''. Due to the
curvature energy in the present universe is too small, $\Omega $$%
_{cu}=-10^{-81}$, it is very difficult to observe its existence. However we
are not able to exclude this possibility of the closed universe, no matter
how we improve on our observation precision.

All just mentioned is high conjecture. We should distinguish the conjecture
of this section with the reasoning in the previous sections. The curvature
energy dependents on the inflaton initial value, which determine finally the
life of a cycle of the universe. Anyhow, the cycle life is finite, but the
number of the cycles may be infinite.

\section{Some conjuncture about the big shrink}

We have owned the evolution rule of the quintessential kinetic energy
shrink, i.e., (\ref{z010920f}) and $\rho _k\propto a^{-6}$. It is
convenience to define the ending time coordinate of the universe, i.e., $%
x\equiv \tau _8-\tau >0$. Then we have $a=a_6(\sqrt{3/2}\dot \phi
_6M_p^{-1}x)^{1/3}$. As a good approximation, we take the balanced point $6$
as the starting point, where the quintessence kinetic energy density is $%
\rho _{k6}\simeq 10^{-364}M_p^4$, the matter density is $\rho _{m6}\simeq
10^{-364}M_p^4$, the radiation density is $\rho _{r6}\simeq 10^{-448}M_p^4$
and the curvature energy density is $\rho _{c6}\simeq -10^{-363}M_p^4$.
Since the quintessence has a large field value $\phi _6\simeq 282M_p$ in
this time, and the quintessence potential energy is not able to increase as
the universe shrinks, we neglect it hereafter.

A question we have to discussed here is whether the matter has transformed
into the radiation. This is possible. Since the proton will decay. However
the baryonic matter is a small part of the whole matter, a large part of
matter is the cold dark matter. The key problem is whether the cold dark
matter will decay, many models think so. If a large part of the cold dark
matter is the neutralino of the supersymmetric particles, and if the $R$
parity is an absolute conserved quantum number, then the neutralino does not
decay$^{[16]}$. To say the least, even if the neutralino decay, its final
states must include the neutrinos, however maybe the neutrinos have a small
non-zero mass. The the final state particles produced by the evaporation of
the black holes are not the pure photons, they include the neutrinos at
least. Therefore it is impossible that all matter can transform into the
radiation. Of course even if the matter transform into the radiation, it
does not affect the qualitative conclusion about the big collapse of the
universe in our model, only change some scenario, for example, the matter
leading attractor is replayed by the radiation leading attractor.

Now we would like to estimate a time point ``$q$'', i.e, ``quantum
gravity'', which will be explained soon. At this time point $q$ the
quintessence kinetic energy density arrive the Planck energy scale, i.e., $%
\rho _{kq}\simeq 1M_p^4$. Note that this is very high density, you note that
even if at the born moment of the universe and the begin of the inflation,
the density is only $10^{-9}M_p^4$ in the last section. It is easy to get $%
a_q\simeq 10^{-61}a_6\simeq 10^{30}a_0\simeq 10^{20}a_u$, $\phi _q\simeq
290M_p$ and $x_q\simeq 1M_p^{-1}$ by using just mentioned formulae in this
section. It is a very strange conclusion, since the energy density of the
universe arrives the Planck energy scale at the Planck time until its
ending, but the size of the universe at that time is larger than our present
universe for $20$ magnitude orders! It is unimaginable. The quantum gravity
has to play its role at this high energy scale in such wide space. How does
this kind of universe with such high energy and such large space scope
shrink further?

We must note that at this time the quintessence mass is very small, but it
is not zero, 
\begin{equation}
\label{z010921b}m_\phi ^2=V^{\prime \prime }\simeq \lambda
^2M_p^{-2}M_0^4\exp (-\lambda \phi /M_p),\qquad \text{for }\phi \gg 0,
\end{equation}
and $m_\phi ^2\simeq 10^{-372}M_p^2$ for $\phi _6$. It is very possible that
the rabid running of the quintessence transforms into its exiting states,
i.e., particles due to its fluctuation. We do not know whether this
transform is thermal. If its temperature is absolutely zero, its evolution
is matter-like $\rho _\phi \propto a^{-3}$. If its temperature is non-zero,
its evolution may be radiation-like $\rho _\phi \propto a^{-4}$ since its
temperature will increase in the compression. We can re-estimate the time at
which the density arrives the Planck scale, and get $a_q\simeq
10^{-30}a_0\simeq 10^{-40}a_u$ for matter-like, and $a_q\simeq a_0\simeq
10^{-10}a_u$ for radiation-like, both have $x_q\simeq 1M_p^{-1}$. Despite of
this Planck scale universe is very smaller than the universe at BBN, but it
is very larger than its born size $a_b\simeq 10^5M_p^{-1}$. According to
this argument we think that the universe never compress to its original
state at its birth. The new cycle must have its new property.

A key problem is when, i.e., at what time, the quintessence kinetic energy
transform to the particles? Maybe at the beginning of the universe big
collapse, maybe at the beginning of the quintessence fast rolling. As
mentioned, this does not affect our main conclusion, the matter dominated
universe is almost same with the matter leading attractor before the
constriction. If the quintessence kinetic energy transforms to matter, the
constriction of the universe is just like the standard shrinkage of the
mater closed universe. Same saying for the radiation-like. It is very
possible that this transform happens at the contractive epoch, since the
quintessence has a smaller mass and the universe at this epoch has very
large deceleration which is far different from the inertia system. This
epoch is a very different from other epochs with small deceleration.

We see that the curvature does not play any key role when begin of
shrinkage, its action is only to change the sign of the Hubble parameter. If
there is a negative cosmological constant, the case is similar. But the
quintessential kinetic energy exists still in the contractive stage even if
for a negative cosmological constant, this is a new phenomena been worth to
pay some attention on it.

\section{Conclusion}

Our model realizes in a relative natural way an ``universe'' which is now
accelerated expanding and will shrink ultimately. There is not the event
horizon problem in our model. It seems that it escapes from the consequence
of Ref.[6]. The way of shrinking is unexpected, i.e., the quintessence
kinetic energy dominated, the velocity of the energy increasing is
unexpectedly fast. Our model is semi-realistic since we obtain a time point
``$u$'' with almost same parameters of the present day universe, specially
the universe age $t_u$ and the acceleration parameter $q_u$. Our model
supply a cycle mechanism to realize the infinitely many universe based on
the possible time cycle evolution, of course it is not excluded the
possibility of the existence of the other unconnected cyclic closed
universes. We do a computer simulation and the analytic analysis, both are
well consistent.

Our model is able to change the whole scenario about the origin and ruin, or
recurrence of the cosmos. Our model begets some interesting problems to be
studied worthily.

An outstanding character of the ``Niagara'' potential is that the
quintessence has a negative mass square in its slow rolling phase, i.e., its
exited particle is a tackyon! We have 
\begin{equation}
\label{z010921c}m_\phi ^2=V^{\prime \prime }\simeq -\lambda
^2M_p^{-2}M_0^4\exp (\lambda \phi /M_p),\qquad \text{for}\qquad \phi <-1, 
\end{equation}
and $m_\phi ^2\simeq -10^{-121}M_p^2$ for $\phi \simeq -2$. Whether this
brings some observable effect is an very interesting problem. Kofman et. al.
research the tackyon effect at the early inflation epoch of the universe$%
^{[17]}$. Maybe this tackyon effect can avoid a problem about the fifth
force as an ordinary long range interaction$^{[18]}$.

Other problem is the following. What is a more exact behavior about the
evolution solution of our model. For example, how to transition exactly at
the interim points $3$ or $6$? How to join exactly the approximation
solutions Eqs.(\ref{z010919a}) and (\ref{z010920a})? How to join exactly the
approximation solutions Eqs.(\ref{z010920b}) and (\ref{z010920f})?

How to realize the rebound of the universe from a quintessential big
collapse to a new big bang, or a new big inflation?

What does the absolute high quintessence kinetic energy in the shrink stage
transform into? Whether the quintessence kinetic energy can transform to
matter or radiation in the contractive stage of the universe? Maybe the
quintessence kinetic energy is also able to transform into a pseudo
cosmological constant if we make a careful design for the quintessential
potential, and then a radical different scenario, a part of deflation, will
appear in the constriction of the universe.

What is the origin of the ``Niagara'' potential? Whether we need to
incorporate the early epoch tracker$^{[10]}$ with it?

Whether we still need the anthropic principle to adjust the theoretical
parameters? We think so.

If the universe will shrink, when it shrink to the size of the extra
dimensions, what happens for the extra dimensions?

I hope one can solve some of these interesting problems in the near future.

We must supply a comprehensive understanding for universe from its birth to
its termination. The unilateral quest of separate scenarios without an
unitive viewpoint is not able for us to arrive a goal to uncover truly the
riddle of our wonderful universe. The essence of the cosmological constant
is various, itsits story is rich.

\bigskip

{\bf Acknowledgment:}

{\hspace*{5mm} This project supported by National Natural Science Foundation
of China under Grant Nos 10047004 and NKBRSF G19990754. The author would
like to thank useful discussions with Profs. J.-S.Chen, Z.-G.Deng and
X.-M.Zhang. }\bigskip
\bigskip

\bigskip
\newpage

{\bf References:}

[1] E.W. Kolb and M.S. Turner, {\sl The Early Universe},

\qquad Addison Wesley (1990).

[2] S.J.Perlmutter et al., Nature 391(1998)51;

\qquad A.G.Riess et al., Astron.J.116(1998)1009.

[3] C.J.Hogan, ``Why the Universe is Just So'', astro-ph/9909295.

[3] A.Linde, {\sl Particle Physics and Inflationary Cosmology},

\qquad Harwood Academic Publishers (1990).

[5] P.de Bernardis et al., Nature 404(2000)955.

[6] S.Hellerman, N.Kaloper and L.Susskind,

\qquad ``String Theory and Quintessence'', hep-th/0104180;

\qquad W.Fischler, A.Kashani-Poor, R.McNees and S.Paban,

\qquad ``The Acceleration of the Universe, a Challenge for String Theory'',

\qquad hep-th/0104181.

[7] B.Ratra and P.J.Peebles, Phys.Rev.D37(1988)3406;

\qquad C.Wetterich, Astron.Astrophys.301(1995)321.

[8] E.Witten, ``The Cosmological Constant From the Viewpoint of String

\qquad Theory'', hep-ph/0002297.

[9] S.Weinberg, Rev.Mod.Phys.61(1989)61.

[10] I.Zlatev, L.Wang and P.J.Steinhardt, Phys.Rev.Lett.82(1999)896;

\qquad P.J.Steinhardt, L.Wang and I.Zlatev, Phys.Rev.D59(1999)123504.

[11] M.S.Turner, ``Time at the Beginning'', astron-ph/0106262.

[12] P.G.Ferreira and M.Joyce, Phys.Rev.D58(1998)023503.

[13] A.H.Guth, Phys.Rept.333(2000)555, see its Eq.(20).

[14] M.Duff, ``State of the Unification Address'', hep-th/0012249.

[15] D.-H.Zhang, Commun.Theor.Phys.35(2001)635, gr-qc/0003079,

\qquad see its Eq.(5).

[16] G.Jungman, M.Kamionkowski and K.Griest, Phys.Rept.267(1996)195.

[17] G.Felder, L.Kofman and A.Linde,

\qquad ``Tackyonic Instability and Dynamics of Spontaneous Symmetry
Breaking'',

\qquad hep-th/0106179.

[18] S.M.Carroll, Phys.Rev.Lett.81(1998)3067.

\bigskip
\bigskip

\newpage

%\end{document}

\begin{figure}[h]
{\epsfxsize=15cm\epsfysize=15cm \centerline{
\epsfbox{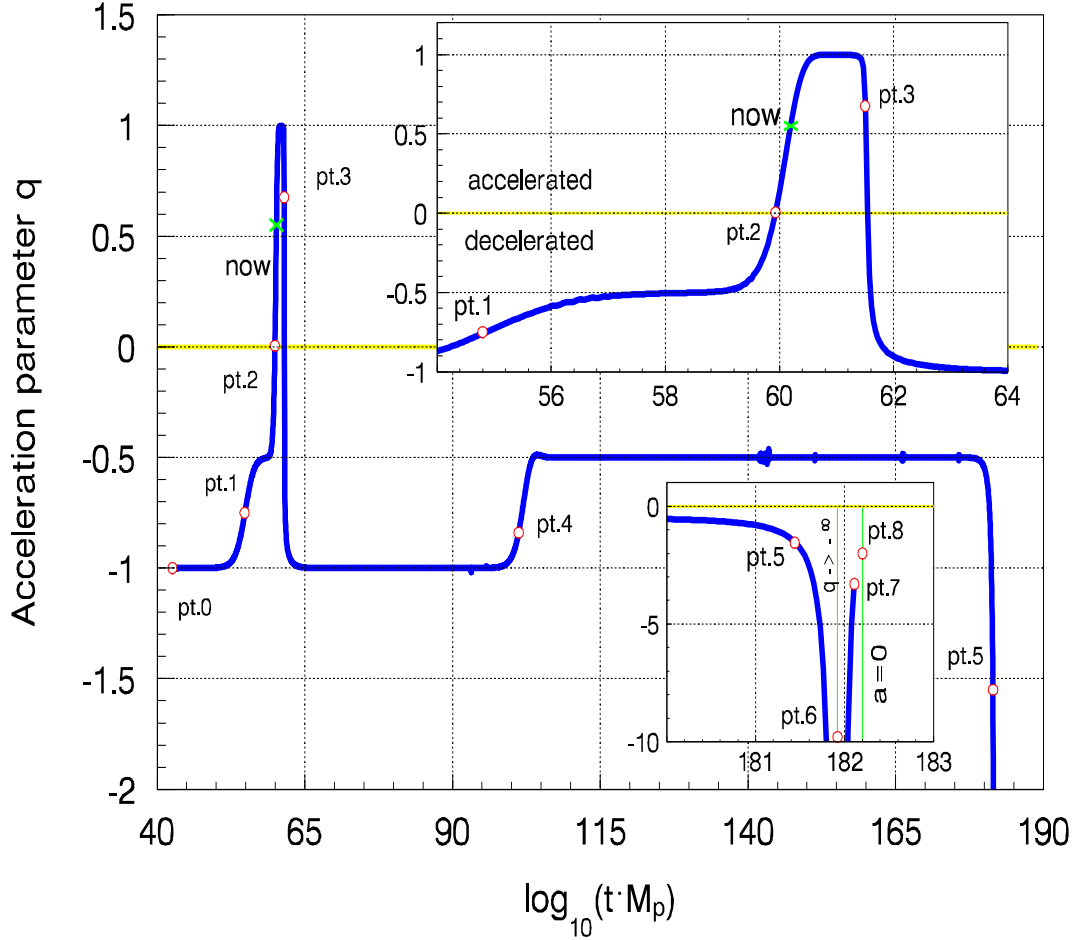}}}
\caption{~~The evolution of the acceleration parameter $q$ 
by computer simulation.
The first stage is the radiation dominated from 
$t_0=0.4\times 10^{43} M_p^{-1}$ to 
$t_1=0.6\times 10^{55} M_p^{-1}$.
The second stage is the matter dominated from $t_1$ to 
$t_2=0.86\times 10^{60} M_p^{-1}$.
The third stage is the quintessence potential energy dominated 
from $t_2$ to 
$t_3=3.2\times 10^{61} M_p^{-1}$.
The fourth stage is the quintessence kinetic energy dominated 
from $t_3$ to 
$t_4=1.6\times 10^{101} M_p^{-1}$.
The fifth stage is the matter leading attractor from $t_4$ to 
$t_5=0.28\times 10^{182} M_p^{-1}$.
The sixth stage is the curvature energy dominated from $t_5$ to 
$t_6=0.85\times 10^{182} M_p^{-1}$.
The seventh stage is a big collapse with the quintessence kinetic energy 
dominated from $t_6$ to 
$t_8=1.74\times 10^{182} M_p^{-1}$. Specially, 
$t_{now}=t_u=1.56\times 10^{60} M_p^{-1}=13Gyr$.
}
\end{figure}

\begin{figure}[h]
{\epsfxsize=15cm\epsfysize=15cm \centerline{
\epsfbox{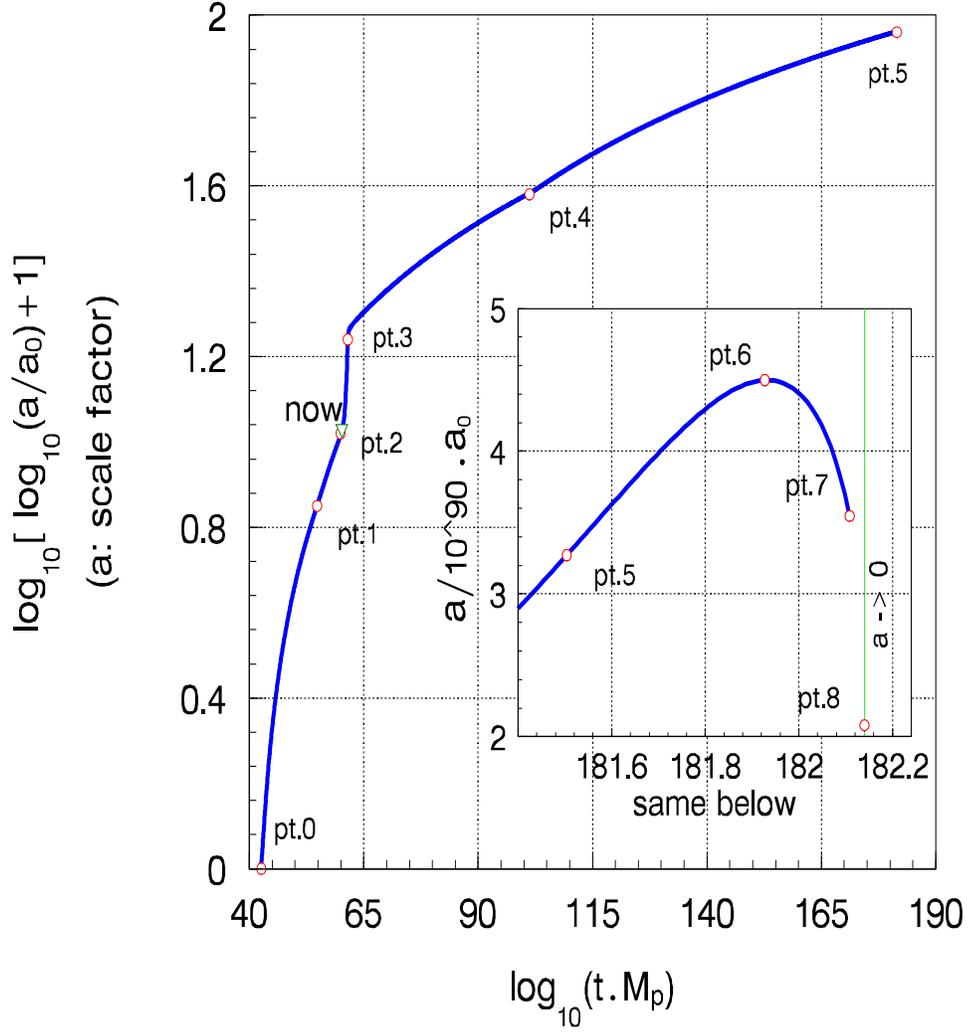}}}
\caption{~~The evolution of the cosmic scale factor $a$.
Because its varying scope is too large, we have to use the double logarithm 
coordinates. In order to get a normalization, we add a minus one in it.
You can see that the scale factor is shrinking in the stage 7.
Some data are, $a_1=1.2\times 10^6 a_0$, 
$a_2=2.4\times 10^9 a_0$, 
$a_{now}=a_u=4\times 10^9 a_0$, 
$a_3=1.7\times 10^{16} a_0$, 
$a_4=1.8\times 10^{37} a_0$, 
$a_5=3.1\times 10^{90} a_0$, 
$a_6=4.5\times 10^{90} a_0$, 
$a_7=0.27\times 10^{90} a_0$, 
and $a_8=0$. 
}
\end{figure}

\begin{figure}[h]
{\epsfxsize=15cm\epsfysize=15cm \centerline{
\epsfbox{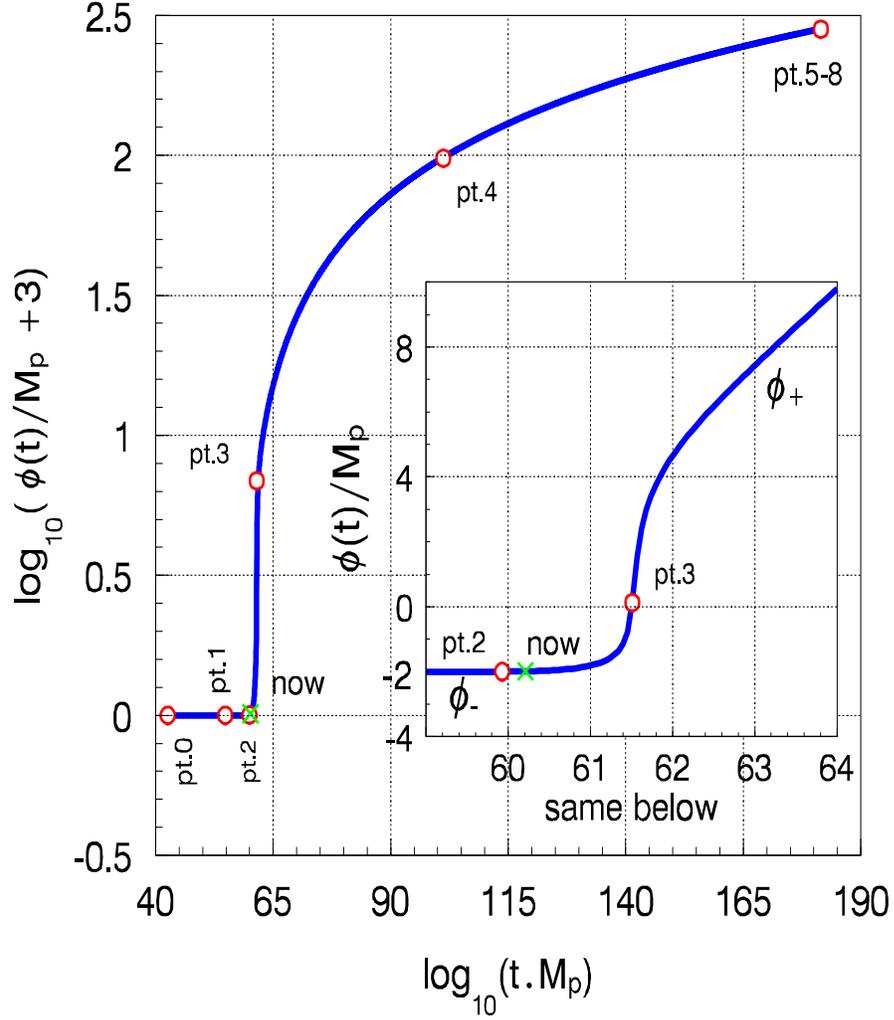}}}
\caption{~~The evolution of the quintessence field value $\phi$.
In order to look at a detail of the behavior in small field value, we use the 
logarithm coordinates. In order to avoid singularity for iniyial value $-2$
, we have to add three in it.
The curve is constructed by $\phi_-$ of Eq.(8) for $\phi<0$ and 
$\phi_+$ of Eq,(12) for $\phi>0$, there is a transition at time $t_3$.
Some data are:
$\phi_0=-2 M_p$,
$\phi_1=-2 M_p$,
$\phi_2=-1.997 M_p$,
$\phi_3=0.026 M_p$,
$\phi_4=95 M_p$,
$\phi_5=280 M_p$,
$\phi_6=281.5 M_p$,
and $\phi_7=289.1 M_p$.
}
\end{figure}

\begin{figure}[h]
{\epsfxsize=15cm\epsfysize=15cm \centerline{
\epsfbox{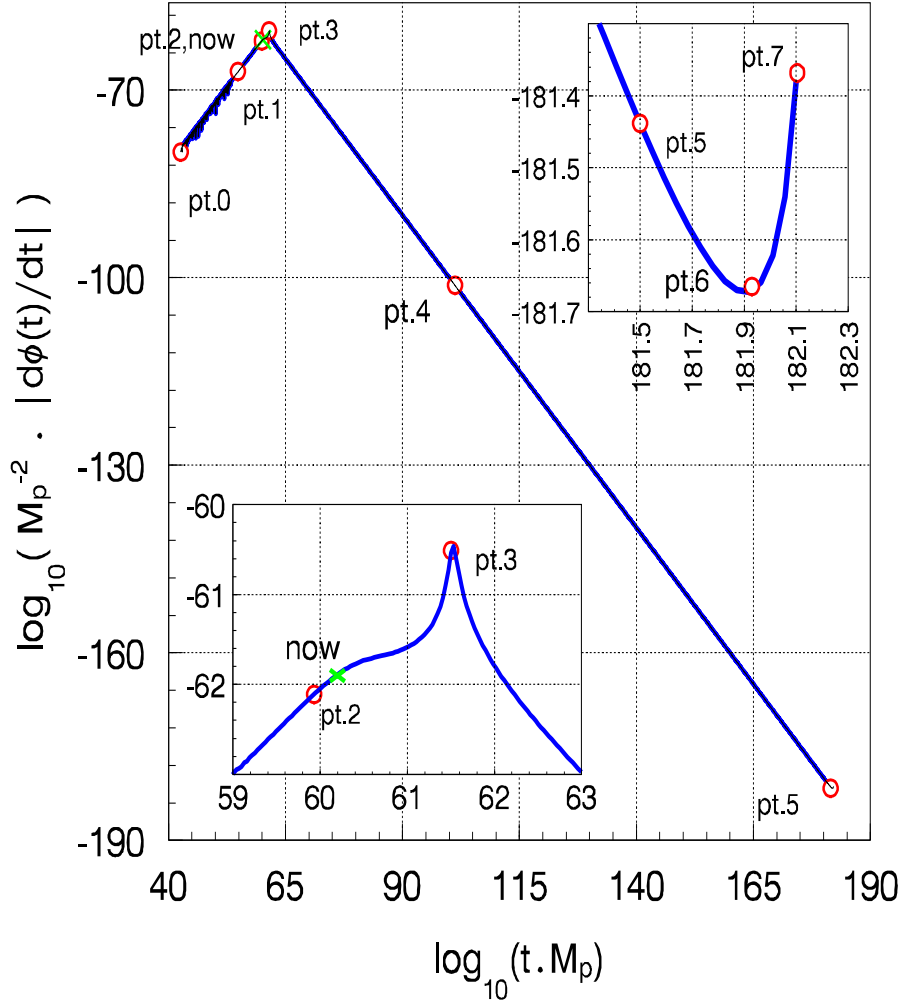}}}
\caption{~~The evolution of the velocity of the quintessence, 
$d\phi /dt$.
This figure includes more information. You can magnify it in your computer 
to watch a detail for the main curve, specially, points $3$ and $5-8$.
Some data are:
$\dot \phi_0=0$,
$\dot \phi_1=0.63\times 10^{-67} M_p^2$,
$\dot \phi_2=0.8\times 10^{-62} M_p^2$,
$\dot \phi_3=0.32\times 10^{-60} M_p^2$,
$\dot \phi_4=0.63\times 10^{-101} M_p^2$,
$\dot \phi_5=0.4\times 10^{-181} M_p^2$,
$\dot \phi_6=0.21\times 10^{-181} M_p^2$,
and $\dot \phi_7=10^{-178} M_p^2$. 
It is notable that the point 6 is not at the minimum of the curve, 
but is at its right side, therefore, $d^2\phi/dt^2>0$, see Eq.(16). 
}
\end{figure}

\end{document}